\documentclass[doublecol]{epl2}
\usepackage{textcomp}
\usepackage{amsmath}
\usepackage{amsfonts}
\usepackage{calc}
\usepackage{mathrsfs}
\usepackage{amssymb}
\usepackage{amsmath}
\usepackage{color}
\usepackage{bm}
\usepackage{graphicx}
\begin{document}
\title{Phonon structures in the electronic density of states of graphene in magnetic field} 
\author{Adam Pound\inst{1} \and J.P. Carbotte\inst{2,3} \and E.J. Nicol\inst{1}}
\shortauthor{Adam Pound \etal} 
\institute{
	\inst{1} Department of Physics, University of Guelph, Guelph, Ontario, Canada, N1G 2W1\\
	\inst{2} Department of Physics and Astronomy, McMaster University, Hamilton, Ontario, Canada, L8S 4M1\\
	\inst{3} The Canadian Institute for Advanced Research, Toronto, Ontario, Canada, M5G 1Z8
}
\date{\today}


\abstract{Unlike in ordinary metals, in graphene, phonon structure can be seen in the quasiparticle electronic density of states, because the latter varies on the scale of the phonon energy. In a magnetic field, quantization into Landau levels creates even more significant variations. We calculate the density of states incorporating electron-phonon coupling in this case and find that the coupling has pronounced new effects: shifting and broadening of Landau levels, creation of new peaks, and splitting of any Landau levels falling near one of the new peaks. Comparing our calculations with  a recent experiment, we find evidence for a phonon with energy similar to but somewhat greater than the optical $E_{2g}$ mode and a coupling corresponding to a mass enhancement parameter $\lambda \simeq 0.07$.}
\maketitle

Nature has presented us with a fascinating solid state
system in the form of graphene. This
material is a two-dimensional crystal of carbon atoms arranged on a 
honeycomb lattice, an apparently simple structure that gives rise to astonishing 
electronic and mechanical properties~\cite{Castro-Neto:09,Abergel:10,Geim:07}.
The charge 
carriers display a linear dispersion at low energies, with the low-energy
limit of the 
tight-binding Hamiltonian mapping onto a relativistic Dirac Hamiltonian
for massless fermions, specifically the Weyl Hamiltonian for massless
neutrinos~\cite{Semenoff:84,Ando:07}. 
In this Hamiltonian, the valence and conduction bands at low energies intersect linearly at a point called the Dirac point (DP). Angle-resolved photemission spectroscopy (ARPES) experiments have observed this band structure~\cite{Bostwick:07}, and correspondingly, scanning tunneling spectroscopy (STS) experiments have measured an electronic density of states (DOS) that grows linearly out of the DP~\cite{Li:09,Miller:09}. However, there is clear evidence that this behavior is altered by many-body renormalizations: in both the ARPES and STS data, the slopes of the linear curves are modified, and kinks or bumps exist at frequencies corresponding to phonon modes~\cite{Zhou:08,Bianchi:10,Bostwick:07,Zhang:08,Li:09, Brar:10} or to electron-hole excitations or electron-plasmon interactions~\cite{Brar:10}. Evidence of many-body renormalizations is also seen in other experiments, such as optical conductivity measurements~\cite{Li:08} that find absorptions beyond the bare prediction~\cite{Gusynin:06,Peres:06}, attributed to electron-electron interactions~\cite{Grushin:09,Peres:10a} and electron-phonon~\cite{Stauber:08,Carbotte:10} interactions that create Holstein sidebands; see ref.~\cite{Peres:10b} for a review.

\begin{figure*}[tb]
\begin{center}
\includegraphics[width=\textwidth]{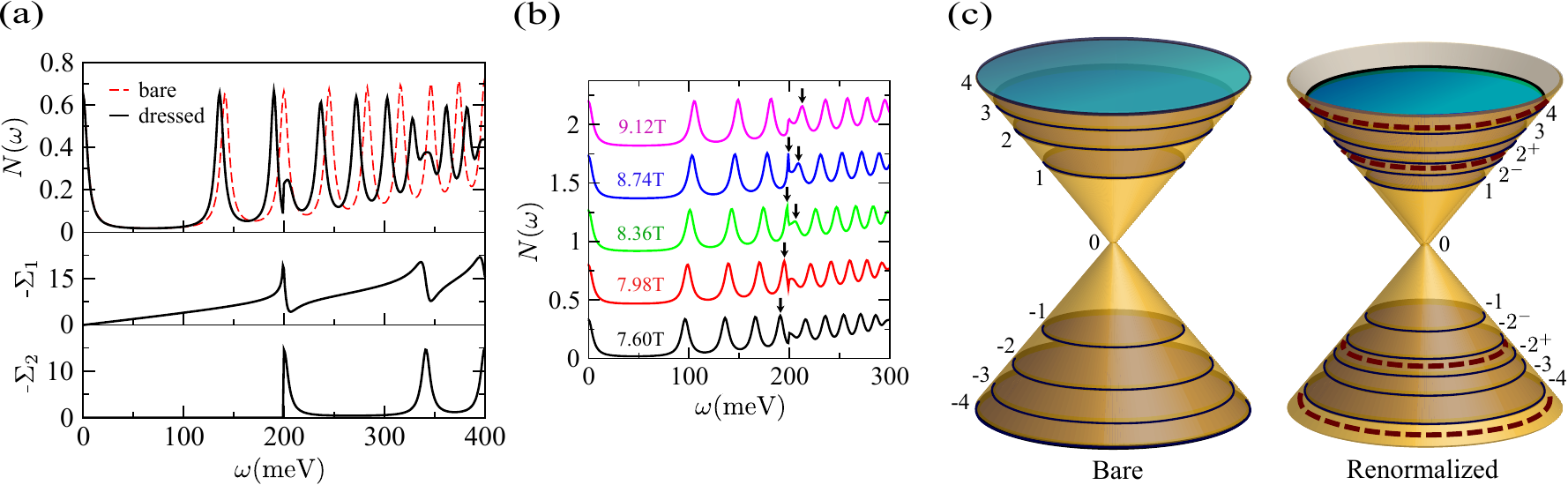}
\caption{(Color online) (a) Bare (dashed red curve) and renormalized (solid black) density of states and self-energy as a function of energy, with $\lambda^{\rm eff}=0.04$, $\omega_E=200$meV, $v_F=10^6$m/s, $\Gamma=5$meV, $\mu=0$, and $B=15$T. For this case of zero chemical potential, the negative-energy region is given by $N(-\omega)=N(\omega)$, $\Sigma_1(-\omega)=-\Sigma_1(\omega)$, and $\Sigma_2(-\omega)=\Sigma_2(\omega)$. The DOS is in units of $N_0$eV, and $\Sigma$ is in units of meV. (b) DOS for increasing values of $B$, with parameters otherwise the same as at left.  Black arrows point to the fourth Landau level, which is split in two when crossing the phonon energy. Curves are offset for clarity. (c) A schematic representation of the position of bare (left) and renormalized (right) energy levels on the Dirac cone. Landau levels are shown as numbered disks outlined in dark blue. Phonon peaks are indicated by the thick dashed red curves. Any Landau level sufficiently close to a phonon peak will be split in two: in this case, $n=\pm2\to\pm2^\pm$.\label{fig1}}
\end{center}
\end{figure*}

In this letter, we focus on one particular many-body effect: phonon signatures in the electronic DOS. In ordinary metals, these signatures are not seen at all, because the DOS is essentially energy-independent. Conversely, in graphene, the DOS varies significantly on the scale of the phonon energy, meaning phonon structure can be seen in it~\cite{Nicol:09,Carbotte:10}.
%The features seen in experiment support this, but the
%evidence is still somewhat limited.
This has allowed scanning tunneling spectroscopy experiments to measure bumps and changes in slope of the DOS (or its first derivative) at energies typical of phonons in graphene, specifically the acoustic out-of-plane mode at 67meV~\cite{Zhang:08,Brar:10} and the A$_{\rm 1g}$ optical mode around 165meV~\cite{Li:09,Brar:10}. (The E$_{\rm 2g}$ optical mode around 200meV, which has been seen in ARPES data~\cite{Bostwick:07}, has not been observed in STS, though a phonon structure of unknown origin has been observed at the somewhat higher value of 240meV~\cite{Brar:10}; this will be discussed below in connection with our own results.) However, evidence still remains somewhat limited.
In a magnetic field~\cite{Sharapov:04,Gusynin:07a,Gusynin:07b} there is much more opportunity to determine the effects of electron-phonon coupling, because the
DOS condenses into Landau levels (LLs), making the energy-dependence of the DOS much more extreme. 
Moreover, the levels can be moved by adjusting the strength
of the magnetic field. In an ordinary two-dimensional electron gas, the
LLs are evenly spaced by the cyclotron frequency $\omega_c=eB/mc$, but in
graphene their energies are proportional to $\sqrt{|n|B}$, where 
$n$ is an integer. In addition, in graphene the $n=0$ LL is field-independent
and fixed to the DP, where there are equal parts hole and electron states~\cite{Li:09,Semenoff:84}. These LLs have been clearly observed in experiments. In particular, tunneling measurements have confirmed the field-dependence~\cite{Li:09,Miller:09,Miller:10,Zeng:10,Song:10}, and optical conductivity measurements~\cite{Jiang:07} have observed the predicted~\cite{Gusynin:07a,Gusynin:07b} transitions between LLs.

Because the magnetic field can moderate the DOS considerably through the LLs,
the amplitude of
phonon structures should be highly dependent
on $B$. Moreover, the quantization of the LLs in graphene in typical
magnetic fields used in experiment is on the scale of the phonon
energy, making it possible to sweep the LLs through the phonon
energy by changing $B$. All of these features have great potential
to identify 
phonon structures, the energy of the
phonon, and the electron-phonon mass renormalization parameter $\lambda$.
In this letter, we characterize the signatures of phonons in the DOS of
graphene in a magnetic field: renormalization of LL energies,
splitting of LLs, phonon structures, and other signatures. We further
investigate these ideas by providing a comparison of the theory
with recent data of Miller et al.~\cite{Miller:09}.

In the absence of a magnetic field, the electronic density of states in graphene varies linearly as $N(\omega)=N_0|\omega+\mu_0|$, where $N_0\equiv\frac{2}{\pi\hbar^2v_F^2}$ and $\mu_0$ is the non-interacting chemical potential. Once a magnetic field $B$ is applied, and accounting for many-body renormalizations
due to an electron-phonon self-energy $\Sigma=\Sigma_1+i\Sigma_2$, the density of states becomes a sum of Lorentzians, one for each Landau level\cite{Sharapov:04}:
\begin{align}
N(\omega)&=N_0\frac{eBv_F^2\hbar}{\pi c}\theta(W-|\omega|)\nonumber\\
&\quad\times\sum_{n=-\infty}^\infty\frac{\Gamma-\Sigma_2(\omega)}{[\omega-\Sigma_1(\omega)+\mu-M_n]^2+[\Gamma-\Sigma_2(\omega)]^2},\label{DOS}
\end{align}
where $M_n={\rm sgn}(n)\sqrt{2|n|eBv_F^2\hbar/c}$ is the energy of the $n$th Landau level, $\mu=\mu_0+\Sigma_1(0)$ is the interacting chemical potential~\cite{Carbotte:10}, $W$ is a high-energy cutoff (taken to be $7$eV throughout this letter), and $\Gamma$ is a residual scattering rate (taken to be constant for simplicity). At zero temperature the self-energy is given by~\cite{Nicol:09,Carbotte:10}
\begin{align}
\Sigma(\omega) &= \frac{1}{W}\int_0^\infty d\nu\alpha^2F(\nu)\int^\infty_{-\infty}d\omega'\frac{N(\omega')}{N_0}\nonumber\\
&\quad\times\left[\frac{\theta(\omega')}{\omega-\nu-\omega'+i0^+}+\frac{\theta(-\omega')}{\omega+\nu-\omega'+i0^+}\right],\label{Sigma}
\end{align}
where $\alpha^2F(\nu)$ is the electron-phonon spectral density. For the moment we assume an Einstein phonon mode at $\omega_E$, described by $\alpha^2F(\nu)=A\delta(\nu-\omega_E)$, where $A$ is the electron-phonon coupling strength. Such a phonon spectrum with $\omega_E=200$meV was shown by Park et al.\cite{Park:07}
to provide a phonon-induced electron self-energy in excellent agreement
with that from full first principle calculations. Note that we do not take into account corrections to
the phonon energy and lifetime provided by the phonon self-energy as
modified by interaction with Dirac electrons. These corrections have been calculated theoretically~\cite{Ando:06,Ando:07}
and their effects have been seen in experiment~\cite{Faugeras:09,Yan:10}. For the most part, the 
effects are small for our purposes.
Raman experiments~\cite{Faugeras:09} have found a 5meV modulation of the energy
of the optical phonon line in magnetic fields up to 30T. In graphite
samples with enhanced
lifetimes of Dirac fermions,
experiments~\cite{Yan:10} find modulations of about 8meV for fields around 5T. Also,
the measurements see splitting of the phonon structure into
three peaks due to
magnetoresonance of the phonon line energy crossing the Landau level. This feature may be relevant for
our discussion when we examine experimental data, and we will return to it then.

Equations \eqref{DOS} and \eqref{Sigma} are solved iteratively, beginning with the bare density of states, obtained by setting $\Sigma\to0$ in eq.~\eqref{DOS}. This zeroth-order approximation $N^{(0)}(\omega)$ has peaks at $\omega=M_n-\mu_0$ for all $n$ (up to the cutoff determined by $W$). The dashed red curve in fig.~\ref{fig1}(a) shows this zeroth-order approximation for zero chemical potential (the case of charge neutrality) and a 15T field, the largest field strength used in tunneling experiments to date~\cite{Li:09,Miller:09,Miller:10,Zeng:10,Song:10}. To determine the locations of the peaks in the renormalized DOS, we evaluate the imaginary part of the first-order approximation $\Sigma^{(1)}$, immediately found to be
\begin{align}
\Sigma^{(1)}_2(\omega)&=-\frac{\pi A}{WN_0}\left[N^{(0)}(\omega-\omega_E)\theta(\omega-\omega_E)\right.\nonumber\\
&\quad\left.+N^{(0)}(\omega+\omega_E)\theta(-\omega-\omega_E)\right].
\end{align}
This carries an image of the DOS, but displaced by $\omega_E$ for $\omega>\omega_E$ and by $-\omega_E$ for $\omega<-\omega_E$. Therefore, the peaks at $\omega=M_n-\mu_0$ in $N^{(0)}(\omega)$ induce peaks at $\omega=M_n-\mu_0\pm\omega_E$ in $\Sigma_2^{(1)}$, as seen in the lower frame of fig.~\ref{fig1}(a). $\Sigma^{(1)}_1(\omega)$ is peaked at these same energies, as shown in the middle frame of fig.~\ref{fig1}(a). The renormalized DOS $N^{(1)}(\omega)$ is obtained by substituting $\Sigma^{(1)}(\omega)$ into eq.~\eqref{DOS}. Since the right-hand side of eq.~\eqref{DOS} contains not just peaks, but a slowly increasing portion contributed by the sums of tails of Lorentzians, peaks in $\Sigma(\omega)$ introduce new peaks in $N(\omega)$ at the same energies. Therefore, after renormalization, $N^{(1)}(\omega)$ has peaks at $\omega=M_n-\mu_0\pm\omega_E$, which we shall refer to as phonon peaks. This is evident in the top frame of fig.~\ref{fig1}(a), where we see phonon peaks at $\omega_E=200$meV and $\omega_E+M_1=340$meV in the renormalized DOS (solid black curve). The magnitude of these peaks is proportional to the electron-phonon coupling strength $A$, to which we have given the small value of $A=50$meV in fig.~\ref{fig1} to make for easy comparison between the bare and renormalized curves.

Along with these new peaks, there are those corresponding to the original Landau levels, now located at solutions to $\omega-\Sigma_1(\omega)+\mu=M_n$. The $n=0$ LL does not shift, remaining pinned at the DP [note that the $n=0$ LL
of the bare DOS is hidden under the black curve in fig. 1(a)]. For small $\omega$, $-\Sigma_1(\omega)\simeq\omega\lambda^{\rm eff}$, where $\lambda^{\rm eff}$ is the effective mass renormalization parameter. As seen in the middle frame of fig.~\ref{fig1}(a), this linear behavior remains a good approximation until $\omega$ approaches the peak at $\omega_E$. For the parameters used in fig.~\ref{fig1}(a), $\lambda^{\rm eff}=0.04$ and the $n=1$ LL is appropriately shifted down by a factor of $1+\lambda^{\rm eff}$, which can be interpreted as a renormalization of the Fermi velocity in $M_n$. But the $n=2$ level, which falls near $\omega_E$, is shifted by a larger factor. Between the first and second phonon peaks, $\Sigma(\omega)$ returns to linear behavior. In addition to their locations, the widths of the LLs are affected by the self-energy: the imaginary part of $\Sigma$ becomes nonzero above $\omega_E$ due to the opening of a new scattering channel, broadening the levels above $\omega_E$ beyond their bare-band widths.

In fig.~\ref{fig1}(b), we show a sequence of curves for the renormalized DOS
at five increasing values of $B$ between 7.6T and 9.12T. The other parameters are identical to those in fig.~\ref{fig1}(a).
For clarity each curve is displaced upwards by 0.35$N_0$eV. This progression
shows what happens when a given LL (indicated by the black arrows) is made to pass through
the phonon energy. In the lower curve (black),
the LL falls below the small phonon peak that begins sharply at $\omega_E$. As we increase
$B$, the LL passes through the phonon peak. In the process, it is split in two, as indicated by the two black arrows, one to each side of $\omega_E$. Because the phonon peak is small for our chosen value of $A$, it is overwhelmed by the right peak in the split LL. At sufficiently high $B$, the left peak vanishes and the right peak moves sufficiently far from $\omega_E$ for the phonon peak to again become visible.

%For both figs.~\ref{fig1}(a) and (b), we have used a field of 15T. While the typical magnetic fields
%used in the tunneling experiments where LLs have been seen have only
%been up to 15T~\cite{Li:09,Miller:09,Miller:10,Zeng:10,Song:10},
%recent work on graphene nanobubbles finds LLs based on strain-induced
%pseudomagnetic fields which can surpass 300T~\cite{Levy:10}.

We illustrate the effects of renormalization on the energy levels in the schematic fig.~\ref{fig1}(c). Comparing the bare LLs (shown on the left Dirac cone) to the renormalized energy levels (on the right Dirac cone), we see the LLs (disks outlined in dark blue) brought to lower energies, the new phonon peaks (dashed red circles) interspersed between them, and LLs split in two (here $n=\pm2\to\pm2^\pm$) when they are sufficiently close to a phonon peak.

\begin{figure}[tb]
\begin{center}
\includegraphics[scale=0.8]{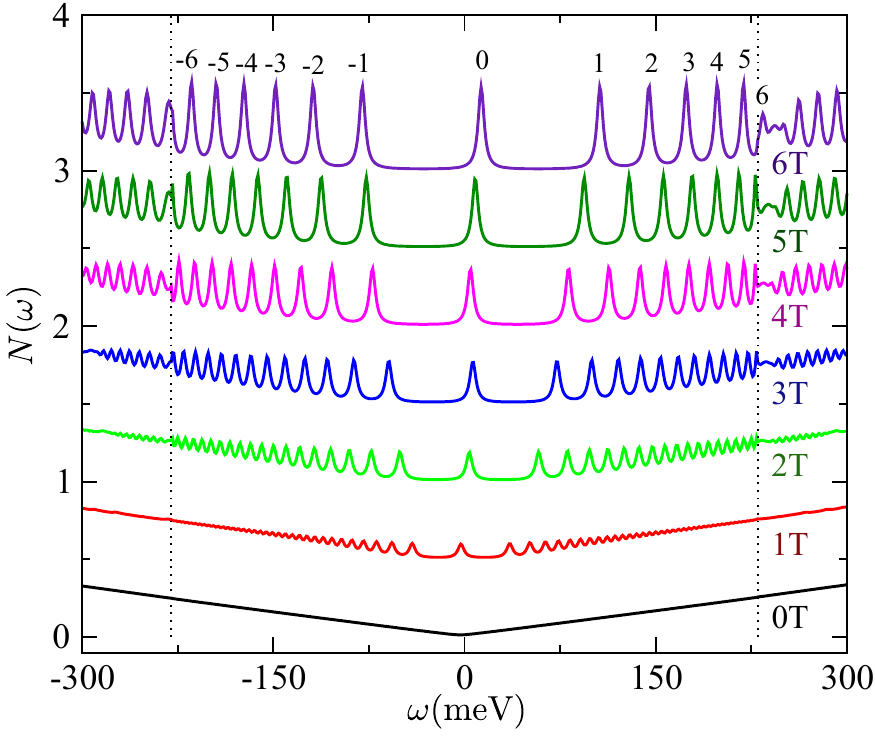}
\caption{(Color online) Density of states as a function of energy for a sequence of magnetic field strengths. Peaks are labeled with their Landau level indices. The dotted lines mark plus and minus the phonon energy. $N$ is in units of $N_0$eV and curves are offset by 0.5$N_0$eV for clarity.\label{fig2}} 
\end{center}
\end{figure}

Our results' experimental relevance is seen by comparing them with the results of Miller et al.\cite{Miller:09}, who obtained DOS curves from scanning tunneling spectroscopy on epitaxial graphene grown on the carbon-face of SiC. In fig.~\ref{fig2} we show
a sequence of theoretical DOS curves calculated for values of magnetic field in the range 0 to 6T, as labelled in the figure. (Curves are again offset for clarity.) Here,  through consideration of the Miller data, we have used parameters 
$\lambda^{\rm eff}=0.07$, $\omega_E=230$meV, $\Gamma=3$meV,
and $v_F=1.13\times 10^6$m/s.
Note that we have also used a finite, though small, chemical potential in order to roughly mimic its increase with $B$ seen by Miller et al.; from bottom to top, $\mu=3.4$meV, 2.8meV, -3.8meV, -6.6meV, -4.8meV, -8.6meV, and -13.3mev.
This shift in $\mu$ with magnetic field has been attributed to a redistribution of charge
in the graphene multilayers, screening, and issues associated with the probe tip~\cite{Miller:09}, and it results in a visible asymmetry between the peak structures at $\omega_E$ and $-\omega_E$.
Clearly, as expected,
the effects of phonon coupling grow along with the increasing amplitude of the LLs as $B$ increases, while for
$B=0$ (continuous black curve) they cannot be seen on the scale of the
figure. As in the preceding discussion, below the phonon energy (indicated by the dotted lines), the LLs
behave as the bare DOS, but 
renormalized by $1+\lambda^{\rm eff}$. Above $\omega_E$, we see
the first phonon peak more prominently as $B$ increases, followed by more
ordinary LLs at higher energy, but with heavily damped amplitudes; these effects are larger than in the results of Fig.~\ref{fig1} due to the larger value of $A$ (or that of $\lambda^{\rm eff}$).
All these features are in qualitative agreement with the data of Miller
et al., who show a similar figure for the same magnetic field values.

\begin{figure}[tb]
\begin{center}
\includegraphics[scale=0.85]{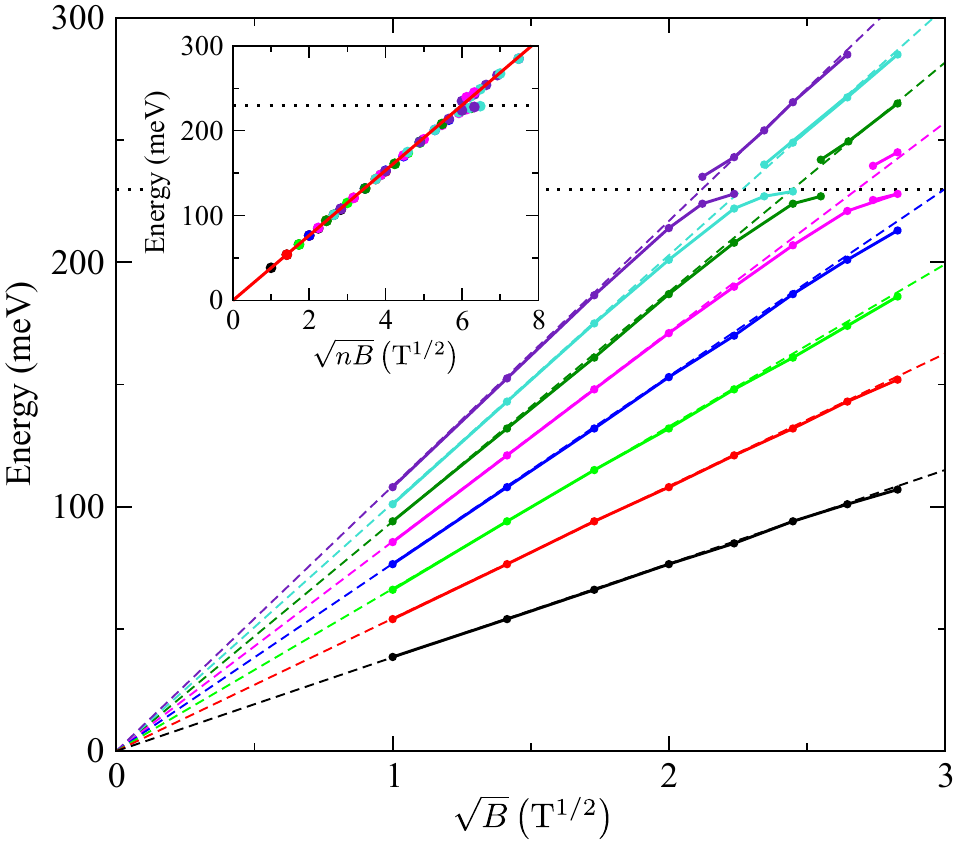}
\caption{(Color online) Locations of peaks in the DOS as a function of $\sqrt{B}$. The circular data points connected by solid curves show the positions of the renormalized Landau levels. The thin dashed lines show $M_n/(1+\lambda^{\rm eff})$. The dotted black line indicates the phonon energy. Curves for negative energies are the mirror image of those shown here. Inset: the same points plotted as a function of $\sqrt{nB}$.\label{fig3}}
\end{center}
\end{figure}

In fig.~\ref{fig3}, we show the energies at which the various peaks fall
(points connected by solid lines to guide the eye) as a function of $\sqrt{B}$. The parameters are as in the preceding figure, but for simplicity, here we use $\mu=0$ for all values of $B$. For small values of
magnetic field, the energy of the $n$th level agrees well with $M_n/(1+\lambda^{\rm eff})$ (dashed lines).
As expected, however, when they approach the phonon energy $\omega_E$ (dotted line), the LLs deviate from this behavior and are split in two about $\omega_E$. In the inset, we show that when plotted as a function of $\sqrt{nB}$, all the curves collapse into a single straight line of slope $v_F\sqrt{2\hbar e/c}$, except for the notable deviation near the phonon energy.

\begin{figure}[tb]
\begin{center}
\includegraphics[scale=0.85]{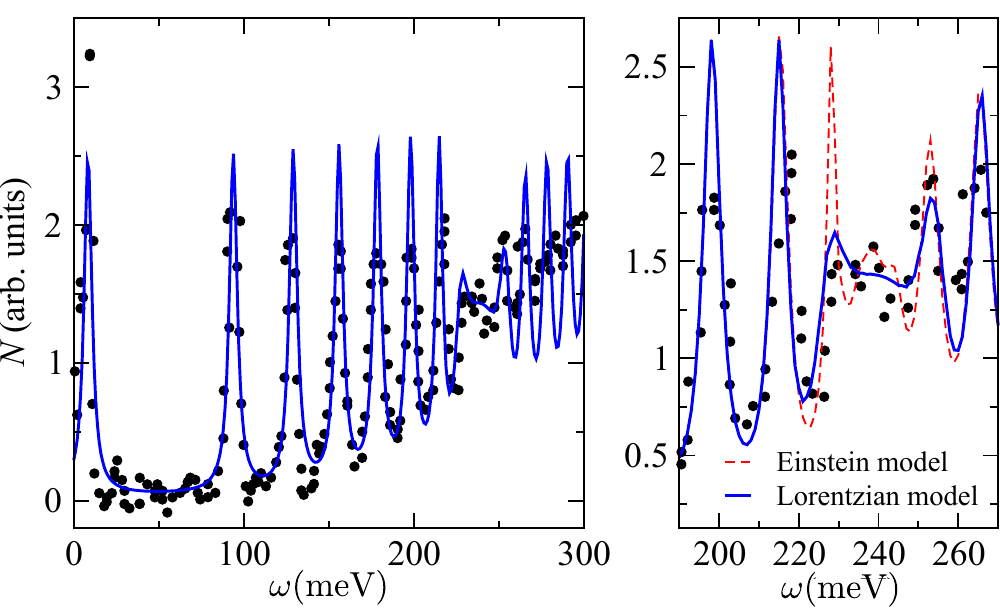}
\caption{(Color online) Density of states as a function of energy at $B=5T$, comparing numerical results to experimental data (solid black circles) taken from Miller et al.\cite{Miller:09}. The plot on the right zooms in on the region around the phonon energy, showing results for both the Einstein model (dashed red curve) and a truncated Lorentzian (solid blue curve).\label{fig4}}
\end{center}
\end{figure}

Figure~\ref{fig4} shows a more detailed comparison with the data (black dots) of Miller et al.\cite{Miller:09} in the case $B=5$T. The solid blue curve is our result
for the renormalized DOS (in arbitrary units) as a function of energy $\omega$,
still with the same parameters as in fig.~\ref{fig2}. We see qualitative and even semiquantitative
agreement (left frame). The phonon structure around 230meV is clearly
seen in the data, as is a clear drop in amplitude of the LLs beyond
this energy. In the right frame, we show an enlarged view of the region around the phonon
energy. Along with results for coupling to a single
Einstein mode (dashed red curve), we show results for the more realistic model of a truncated Lorentzian phonon distribution, centered
at 230meV with a half-width of $10$meV and a cutoff at $|\omega-\omega_E|=20$meV. In the former case, the $n=7$ Landau level is sharply split around the phonon peak; in the latter case, the phonon peak is broadened and merges with the split Landau level, creating the flattened structure seen in the data. This structure lies at an energy roughly 30meV greater than that of the $E_{\rm 2g}$ mode. A phonon structure at a similar energy, 240meV, was also identified by Brar et al.~\cite{Brar:10} in their STS study of graphene flakes on SiO$_2$ with no magnetic field, and they suggest that it could be due to a multi-phonon process. Alternatively, it might be the $E_{\rm 2g}$ mode shifted to higher frequency by some other mechanism. For example, tip effects such as deformation of the sample could cause a shift~\cite{Thomsen:02}, and other anomalous features in Miller's data have been ascribed to them~\cite{Kubista:11}. A shift could also be caused by the magnetoresonance effect seen in Raman measurements\cite{Faugeras:09,Yan:10}, which is especially large at 5T. Regardless of the origin of this feature, our model provides a reasonable fit to the data and gives a rather small value of effective mass renormalization of $\lambda^{\rm eff}\simeq 0.07$.

In conclusion, 
we have calculated the effect of electron-phonon interactions on the
density of states $N(\omega)$ in graphene in a magnetic field $B$. The imaginary part
of the electron-phonon self-energy gives the electron scattering rate associated with an electron scattering between states of energy $\omega$ and
$\omega-\omega_E$, where $\omega_E$ is the phonon energy. This
process is possible only for $\omega>\omega_E$, and it carries an
electronic density of states factor $N(\omega-\omega_E)$, encoding information on the underlying LL structure. From this encoding, the electron-phonon coupling  generates a new set of peaks, each one an image of a bare Landau level displaced by the phonon energy. As the
magnetic field is increased, both the LL peaks and the 
new peaks are increased in amplitude. At energies small compared to $\omega_E$,
the renormalized Landau levels closely follow $E_n={\rm sgn}(n)\sqrt{2|n|eBv_F^2\hbar/c}/(1+\lambda^{\rm eff})$, where $\lambda^{\rm eff}$ is the electron-phonon mass renormalization parameter. As $\omega_E$ is approached, however, the LL is shifted beyond the
$1/(1+\lambda^{\rm eff})$ renormalization factor; furthermore, it is split in two about the new peak at $\omega_E$. Since for $\omega>\omega_E$
a new scattering channel opens, the resulting
amplitudes of the LL lines above $\omega_E$ are also reduced over their bare band value and 
their widths increased. All these features are seen in the STS data. Fitting to this data at $B=5$T provides a new estimate of the electron-phonon mass enhancement $\lambda^{\rm eff}\simeq0.07$.

\begin{acknowledgements}
This work was supported in part by the Natural Sciences and Engineering Research Council of Canada and the Canadian Institute for Advanced Research.
\end{acknowledgements}

\bibliography{DOS_short}

\end{document}